\pdfoutput=1

\documentclass{iaus}
\usepackage{graphicx}

\usepackage{epsf}
\usepackage{epstopdf}

\title[Current helicity constraints in solar dynamo models] 
{Current helicity constraints in solar dynamo models}

\author[D. Sokoloff et al.]   
{D.Sokoloff$^1$,
H. Zhang$^2$,
D. Moss$^3$,
N.Kleeorin$^4$,
K.Kuzanyan$^5$,
I.Rogachevski$^6$,
Yu Gao$^2$
\and H. Xu$^2$}

\affiliation{$^1$Department of Physics, Moscow State University, Moscow
119992, Russia \\ email: {\tt sokoloff.dd@gmail.com} \\[\affilskip]
$^2$National Astronomical Observatories, Chinese Academy of
Sciences, Beijing 100012, China \\ email: {\tt hzhang@bao.ac.cn}\\[\affilskip]
$^3$School of Mathematics, University of Manchester,
Manchester M13 9PL, UK  \\ email: {\tt moss@maths.man.ac.uk}\\[\affilskip]
$^4$Department of Mechanical Engineering, Ben-Gurion University
of Negev, POB 653, 84105 Beer-Sheva, Israel \\
email: {\tt nat@menix.bgu.ac.il} \\[\affilskip]
$^5$IZMIRAN, Troitsk, Moscow Region 142190, Russia \\
email: {\tt kuzanyan@izmiran.ru} \\[\affilskip]
$^4$Department of Mechanical Engineering, Ben-Gurion University
of Negev, POB 653, 84105 Beer-Sheva, Israel \\
email: {\tt gary@bgu.ac.il} \\[\affilskip]
}

\pubyear{2012}
\volume{294}  
\pagerange{119--126}
\jname{Solar and Astrophysical Dynamos and Magnetic Activity}
\editors{A.G. Kosovichev, E.M. de Gouveia Dal Pino \& Y.Yan, eds.}
\begin{document}

\maketitle

\begin{abstract}
We investigate to what extent the current helicity distribution observed in solar active regions is compatible with solar dynamo models. We use an advanced 2D mean-field dynamo model with dynamo action largely concentrated near the bottom of the convective zone, and dynamo saturation based on the evolution of the magnetic helicity and algebraic quenching. For comparison, we also studied a more basic 2D mean-field dynamo model with simple algebraic alpha quenching only. Using these numerical models we obtain butterfly diagrams\textbf{ for both} the small-scale current helicity and the large-scale magnetic helicity, and compare them with the butterfly diagram for the current helicity in active regions obtained from observations. This comparison shows that the current helicity of active regions, as estimated by $-{\bf A \cdot B}$ evaluated at the depth from which the active region arises, resembles the observational data much better than the small-scale current helicity calculated directly from the helicity evolution equation. Here ${\bf B}$ and ${\bf A}$ are respectively the dynamo generated mean magnetic field and its vector potential.
\end{abstract}

\firstsection 

\section{Introduction}

The solar activity cycle is believed to be a manifestation of dynamo
action somewhere in solar interior which generates waves of quasi-stationary magnetic field
propagating from middle latitudes towards the solar equator (\textbf{``}dynamo
waves"). The traditional explanation of
this dynamo action (Parker, 1955) is based on  the joint action of
differential rotation and mirror asymmetric convection which results
in what has come to be known as the $\alpha$-effect,
based on the helicity of the
hydrodynamic convective flow. This explanation is however not the
only one currently discussed in the literature
and, for example,  meridional circulation is also
suggested as an important co-factor of the $\alpha$-effect, see
{
e.g., Dikpati and Gilman, (2001); Choudhuri et al., (2004).}

In turn, traditional dynamo scenarios based on differential rotation
and the classical $\alpha$-effect have to include a dynamo
saturation mechanism. One of the most popular saturation mechanisms
is based on a contribution to the $\alpha$-effect from magnetic fluctuations  (
{Pouquet} et al., 1976). A relevant quantification of this
effect involves considerations of magnetic helicity evolution(e.g., Kleeorin et al., 1995; 2003). Again, this scenario is not the
only one that has been suggested: for example 
{Brandenburg (2007}) considers coronal-mass ejections as an important part of nonlinear suppression of the dynamo, and 
{Mitra et al. (2011)} consider the effects of the solar wind.

A natural way to resolve such controversies is to determine relevant
quantities such as the $\alpha$-effect through observations,
thus, providing a check on the various scenarios.
Such an option is now becoming
realistic, starting from the 1990s when the first attempts to
observe current helicity in solar active regions have been undertaken (Seehafer, 1990; Pevtsov et al., 1994;
Bao and Zhang, 1998; Hagino and Sakurai, 2004).

Twenty years of continuous efforts by several observational groups,
with the most 
{systematic} contribution coming from the Huairou Solar 
{Observing} Station of
China, have resulted (Zhang et al., 2010) in reconstruction of the current
helicity time-latitude (butterfly) diagram for one full solar
magnetic cycle (1988--2005). From this butterfly diagram it is
apparent that the current helicity is involved in the solar activity
cycle and follows a polarity law comparable with the Hale polarity
law for sunspots -- but rather more complicated. In other words,
the dynamo generated magnetic field is indeed mirror asymmetric, and this
mirror asymmetry is involved in the solar activity cycle, and can be
used to understand its nature (Kleeorin et al., 2003, Zhang et al., 2006).

The natural step now is to compare the observed current helicity butterfly diagrams with predictions of particular
dynamo models of the solar activity cycle. We performed such a comparison for a dynamo model  in which
dynamo action is more effective in the deeper layers of the solar convective zone simply because the effect is just from the $\Omega$ gradients being larger there (Zhang et al., 2012). Here we summarize our findings in the context of solar dynamos avoiding technical details.

\section{The observed current helicity butterfly diagram and dynamo models}

The general structure of the observed current helicity butterfly diagram
for the last two solar cycles can be described
as follows. Current helicity is involved in the solar activity cycle
and follows a polarity rule comparable to (however, more complicated than) the polarity rule for toroidal magnetic field, which in
turn comes from the Hale polarity rule for sunspot groups.
Migration of the helicity pattern is clearly visible and located
near the toroidal field pattern. The wings of the helicity
butterflies are slightly more inclined to the equator than the
magnetic field wings, but the former follow in general the latter.
{Though the current helicity as a quantity is not strictly speaking quadratic in magnetic field, in one the same hemisphere it has has the same predominant sign for both consequently observed solar cycles,
}
 with the opposite sign in the other hemisphere (a kind of unchanging dipolar symmetry). There are some domains in the diagram where the current
helicity has the reversed sign with respect to the global polarity law.
These domains of "wrong" sign are located at the very beginning
and the very end of the wings: see Figure 1 (also Figure 1 of 
{Zhang et al. (2012). This regularity has been observed in available data for both solar cycles 22 and 23}.

\begin{figure}
\begin{center}
\includegraphics[height=0.86\textwidth,angle=90]{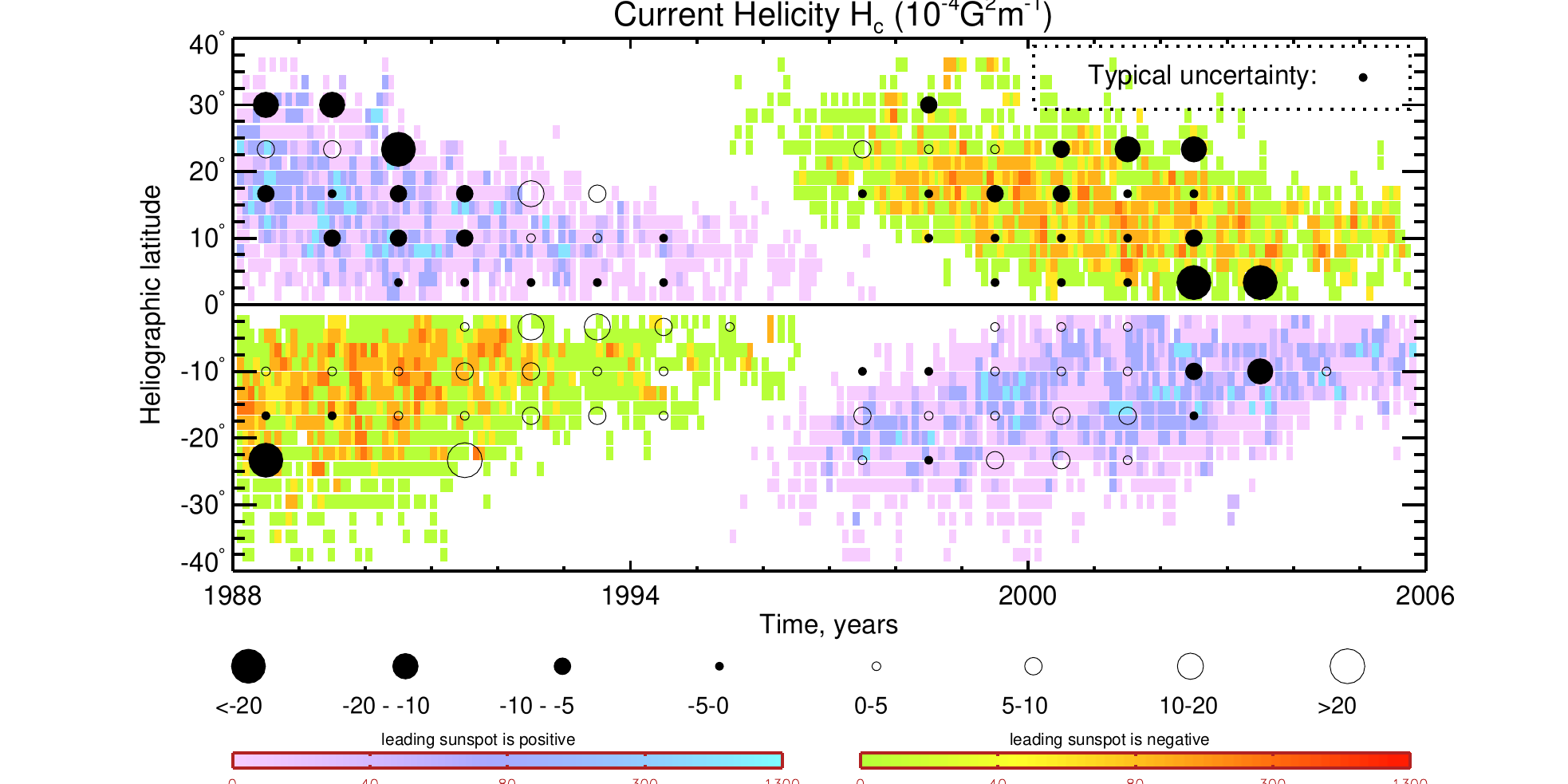}
\caption{Observed current helicity (white/black circles for
positive/negative values) for solar active regions in the 22$^{nd}$
and 23$^{rd}$ solar cycles as averaged over two-year running windows
over latitudinal bins of 7 degrees wide, overlaid with sunspot
density (colour). The circle in the upper right corner of the panel
indicates the typical value of observational uncertainty defined by
95\% confidence intervals scaled to the same units as the circles.
72 out of 88 groups (82\%) have error bars that are smaller than the
signal level. The vertical axis gives the latitude in degrees and
the horizontal gives the time in years. 
Taken from Figure 1 in Zhang et al. (2012).} 
\label{Fig0}
\end{center}
\end{figure}

Our approach to compare the dynamo models with observations is as
follows. We consider two types of dynamo models. 
Both types of models
are 2D mean-field models with an axisymmetric magnetic field that
depends on solar radius $r$ and polar angle $\theta$.
The third (azimuthal) coordinate is $\phi$, and $\partial/\partial\phi=0$.
The dynamo action is based on differential rotation, with a
rotation curve which resembles that of the solar convection zone,
as known from
helioseismological observations, and there is a conventional $\alpha$-effect.

The first type of models assumes a very naive algebraic
$\alpha$-quenching. Then we suppose that the total
magnetic helicity is locally vanishing, so the magnetic helicity of the
large-scale magnetic field produced in the  course of mean-field
dynamo action has to be compensated by small-scale magnetic
helicity. (Thus, we assume that at initial instant
the medium is non-magnetic, so that helicity conservation means that the sum of large and small-scale helicities
remains locally zero.)
We assume also that there is a separation of scales,
so that characteristic turbulence scales are much smaller
than the characteristic spatial scales of mean magnetic field variations. 
{This allows us to link current
and magnetic helicities.
}
This concept underlies the observational procedure
for determining the current helicity of active regions,
and for calculating the
current helicity from the magnetic helicity
of the small-scale fields. Based on the same
concept we estimate the
large-scale magnetic helicity as $B_\phi A_\phi$,
where $A(r,\theta)\hat\phi$ is the magnetic potential for the poloidal
field. As a result we obtain (for a given radius $r$) a theoretical
model for the current helicity as a function of $t$ and $\theta$
which we overlay on the butterfly diagram for $B_\phi$. We compare
the result with the current helicity butterfly diagram known from
observations and obtained using similar underlying concepts.

We do not consider this primitive scheme as absolutely realistic.
We are sure that any more or less realistic scenario for solar
dynamo suppression have to be much more sophisticated. On the other
hand, we see that this primitive model produces a
helicity butterfly diagram that is quite similar to that observed.
The only shortcoming of the model is
that the maximum current helicity occurs later than the maximum of
$B_\phi$, while it is observed to occur probably earlier, 
{though the amount of data available is probably too little to strongly support this statement}. 

{
The second type of dynamo models with dynamic suppression of the $\alpha-$~effect is based on helicity helicity conservation principle.
}
If magnetic
helicity conservation determines the nonlinear dynamo suppression, we expect
that a careful reproduction of this balance, including helicity fluxes
and the link between magnetic helicity and $\alpha$-effect, will
result in even a better theoretical butterfly diagram, and possibly
improve the phase relations between helicity and toroidal magnetic
field. 

As a specific model that takes into account the influence of
magnetic helicity balance on dynamo action we use the dynamo model
with dynamo action occurring most strongly near the bottom of the convective zone
(the model is described in detail by 
{Zhang et al., (2006)}). Whereas simple $\alpha$-quenching provides a
quite robust suppression of a spherical dynamo and gives (more-or-less)
steady nonlinear magnetic field oscillations for a very wide
range of parameters, in contrast it is far from clear {\it a priori}
that a dynamo suppression based on magnetic helicity conservation is
effective enough to suppress magnetic field growth and result in
steady oscillations. In fact it works more-or-less satisfactorily
only in a quite narrow parameter range, which
appears inadequate to fit observations  convincingly.

We note two crucial points here. First of all, both types of  models
ignore any direct action of magnetic force on the rotation law.
In the more primitive models, there is a crude parametrization
of feedback onto the (purely hydrodynamic) alpha effect.
The latter,  formally more sophisticated model, describes the back-reaction of the
generated magnetic field on the dynamo process in terms
of the magnetic contribution of the current
helicity onto the magnetic part of the $\alpha$-effect. On the other hand,
the feedback of the generated large-scale magnetic field on
turbulent convection is described in our model by the algebraic
quenching of the $\alpha$-effect, turbulent pumping and turbulent
magnetic diffusion.

We assume that helicity conservation is
not the only mechanism of dynamo suppression. The fact that we
see a manifestation of helicity on the solar surface tells us that the
buoyancy must play some role, and we add it to the model. We
stress that the buoyancy which we include in the model transports
current helicity and magnetic helicity as well as large-scale
magnetic field.

\section{Results}

We performed an extensive numerical investigation of the models
in a parameter range considered to be relevant to solar
dynamos.

For the primitive model we found that the simulated current helicity butterfly diagram the plots successfully represent the main features of the
observed helicity patterns.
 Of course, it is possible to choose a set of dynamo governing
parameters which may be less similar to those used to describe the Sun. For example
the role of magnetic fields in the deeper layers of the convective
zone (say, in the overshoot layer) can be emphasized by
adjusting the profile of turbulent diffusivity.
This tends to make the helicity wave in the
overshoot layer look more like a standing wave, but however keeps
the main features of the surface diagram. The highly
anharmonic standing patterns of butterfly diagrams that were discussed
as a possible option for some stars, see 
{Baliunas et al., (2006)}, look however
to be irrelevant for the solar case.

Of course, the helicity pattern in the butterfly diagram obtained in the
models for particular choices of dynamo governing parameters can be
slightly different from the observed helicity patterns.
{Xu et al., (2009)} demonstrated that meridional circulation can be used
to make the simulated pattern resemble more closely to what observed (see Figure 2 in their paper).

\begin{figure}
\begin{center}
\includegraphics[width=0.76\textwidth,height=0.30\textwidth,angle=0]{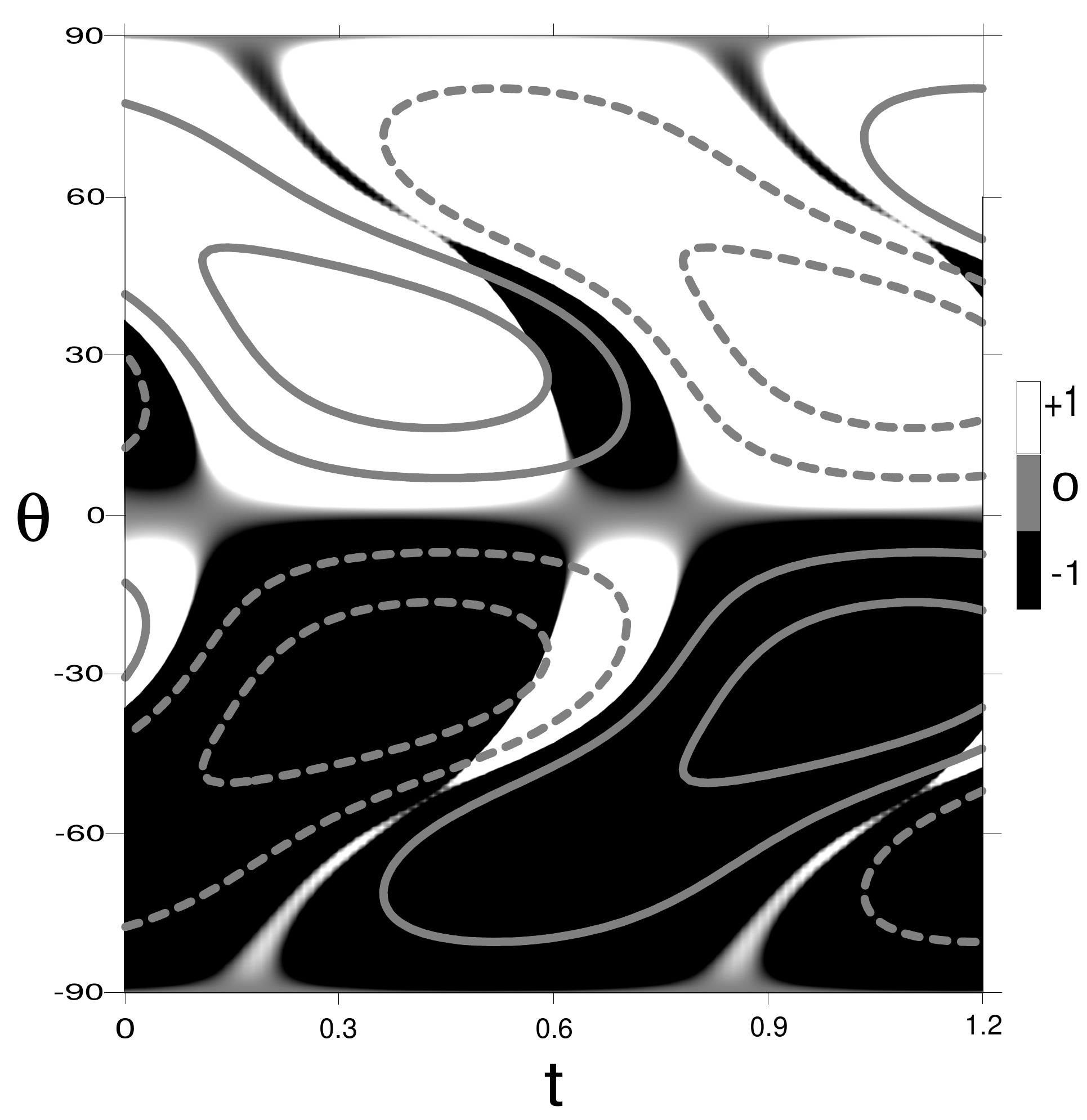}
\caption{Simulation of butterfly diagram for current helicity (greyscale) over the amplitude of the toroidal magnetic field (contour plot) in simple model of Parker dynamo wave with algebraic $\alpha-$~quenching. From Xu et al. (2009).} 
\label{Fig2}
\end{center}
\end{figure}

Zhang et al. (2012) in numerical simulation produced the same type of plots for models based on helicity
conservation, to learn that the more or less pronounced migrating pattern is associated with the large-scale
magnetic helicity only. Small-scale current helicity displays only relatively weak vacillatory behaviour, see Figure~5 in Zhang et al. (2012).
We see that the small-scale current helicity is strongly concentrated in middle latitudes and
helicity oscillations which are available in the model are almost invisible on the background of the intensive belt of constant helicity in middle latitudes. We doubt that such oscillations would be observable.
We stress that, if this model produces any travelling helicity pattern,
it is situated in the deep layers only.

The simple models that we have considered here were not intended
to reproduce very fine details of the spatial-time distribution of
helicity observable in the form of butterfly diagrams,
rather to investigate more generally the possible trends and possibilities.
We again note that our modelling of helicity in the solar
convective zone and active regions is still too simplified to 
{be able to} detect
more detailed properties of the helicity dynamics.

\section{Conclusions}

We conclude that, at least in the framework of traditional models of the solar dynamo based on dynamo action associated predominately with the deeper regions of the solar convective zone, the current helicity of the  magnetic field in active regions
is a tracer of the magnetic helicity of  the large-scale magnetic field
in the solar interior. We believe that this provides a unique option
for tracing this quantity, which is very important for the solar dynamo.
According to the observational data (Zhang et al., 2010), the  current
helicity in active regions is mainly negative in the Northern
hemisphere. Numerical models give a negative value for
$\bf - A \cdot B$ in the surface layer of the convective zone
in the  Northern hemisphere, see Figures 3 and 4 of Zhang et al (2012).

\begin{figure}
\begin{center}
\includegraphics[width=0.76\textwidth,height=0.32\textwidth,angle=0]{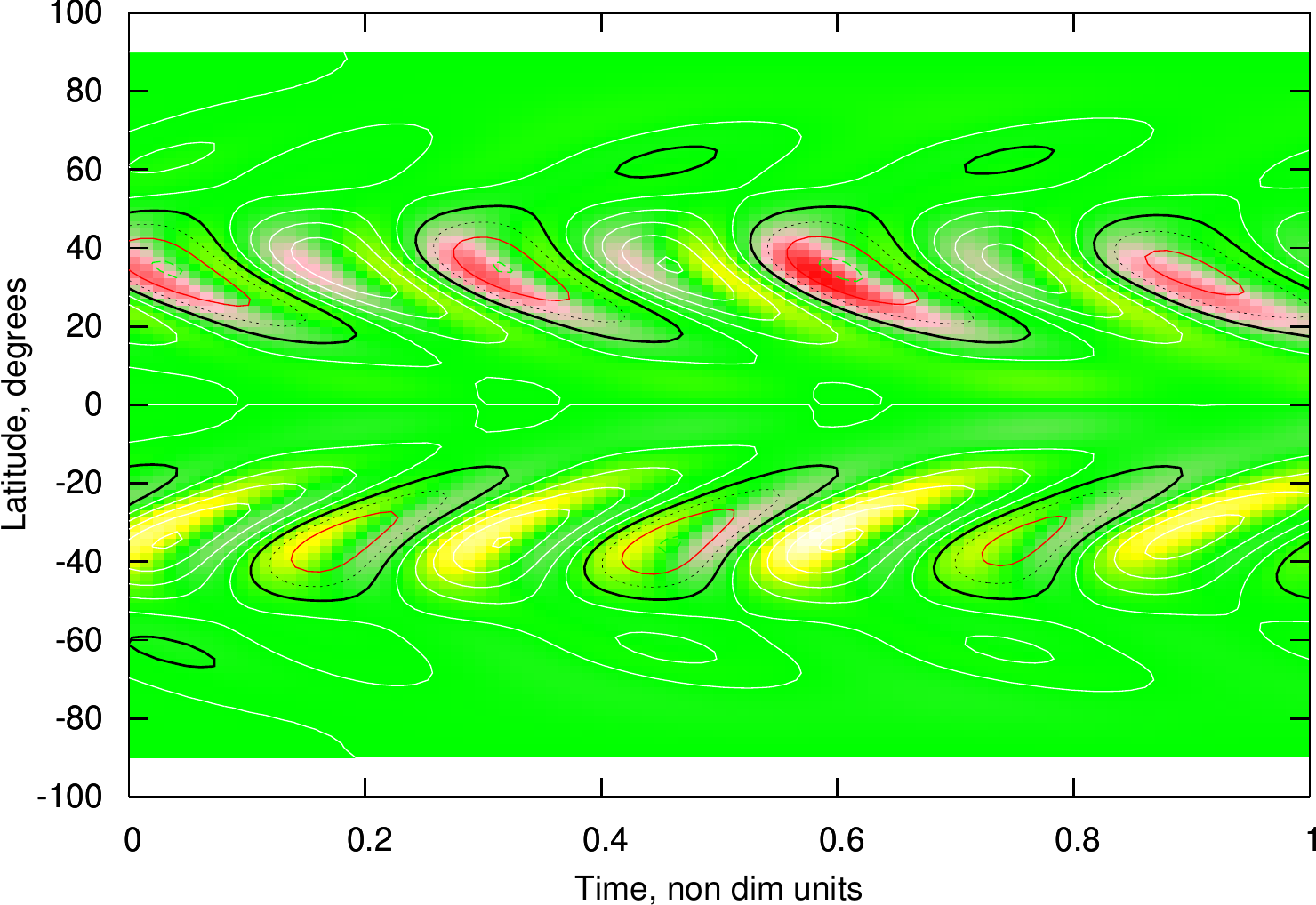}
\caption{Simulation of butterfly diagram for current helicity (colour palette: yellow is positive, red is negative, and  green is zero) over the amplitude of the toroidal magnetic field (contour plot) at the solar surface for the dynamo model with the overshoot layer. Adopted from Figure 3b of Zhang et al. (2012).} 
\label{Fig4}
\end{center}
\end{figure}

Summarizing, we conclude that the current helicity of the magnetic
field in active regions is expected to have the opposite sign to
$\bf A \cdot B$, evaluated at the depth at
which the active region originates.
Thus, the models presented here are consistent with the
interpretation that the mechanism responsible for the sign of the
observed helicity operates near the solar surface.
The mechanism of formation of the current helicity in active
regions still requires further investigation.

\vspace{1mm}

%
D.S and K.K. would like to acknowledge support from Visiting Professorship Programme of Chinese Academy or Sciences 2009J2-12 and NAOC of CAS for hospitality, as well NNSF-RFBR grant 08-02-92211 and Russian grant RFBR 10-02-00960a.
D.S. is thankful to the  RFBR grant 12-02-00170a.
{ H.Z., Y.G. and H.X. are thankful to the Chinese grants 10921303, 11103037, 41174153 and Chinese Academy of Sciences under grant KJCX2-EW-T07.}

\end{document}